\documentclass[showpacs,prd,reprint,preprintnumbers,amsmath,amssymb,superscriptaddress,nofootinbib]{revtex4-1}

\usepackage{graphicx,subfigure,multirow}
 \usepackage{longtable}
\usepackage{dcolumn}
\usepackage{bm}
\usepackage[all]{xy}
\usepackage{color}
\usepackage{slashed}
\usepackage[usenames,dvipsnames]{xcolor}
\usepackage[unicode=true,pdfusetitle,
 bookmarks=true,bookmarksnumbered=false,bookmarksopen=false,citecolor=Turquoise,
 breaklinks=false,pdfborder={0 0 1},backref=false,colorlinks=true]
 {hyperref}
 
 \usepackage[normalem]{ulem}

\usepackage{ulem}

\makeatletter

\newcommand{\fmslash}[2][0mu]{%
  \mathchoice
    {\fmsl@sh\displaystyle{#1}{#2}}%
    {\fmsl@sh\textstyle{#1}{#2}}%
    {\fmsl@sh\scriptstyle{#1}{#2}}%
    {\fmsl@sh\scriptscriptstyle{#1}{#2}}}
\newcommand{\fmsl@sh}[3]{%
  \m@th\ooalign{$\hfil#1\mkern#2/\hfil$\crcr$#1#3$}}
\makeatother

\newcommand{\lsim}{{\;\raise0.3ex\hbox{$<$\kern-0.75em\raise-1.1ex\hbox{$\sim$}}\;}}
\newcommand{\gsim}{{\;\raise0.3ex\hbox{$>$\kern-0.75em\raise-1.1ex\hbox{$\sim$}}\;}}

\newcommand{\beq}{\begin{equation}}
\newcommand{\eeq}{\end{equation}}
\newcommand{\bea}{\begin{eqnarray}}
\newcommand{\eea}{\end{eqnarray}}
\mathchardef\minus="002D
\newcommand{\dis}[1]{\begin{equation}\begin{split}#1\end{split}\end{equation}}

\usepackage{color}

\begin{document}
\title{Supersymmetric Clockwork Axion Model and Axino Dark Matter}
\author{Kyu Jung Bae}
\email{kyujung.bae@knu.ac.kr}
\affiliation{
Department of Physics, Kyungpook National University, Daegu 41566, Korea}
\author{Sang Hui Im} 
\email{imsanghui@ibs.re.kr}
\affiliation{Center for Theoretical Physics of the Universe,
Institute for Basic Science (IBS), Daejeon, 34126, Korea}

\preprint{CTPU-PTC-20-08
}

\begin{abstract}
{
Implications of supersymmetrizing the clockwork axions are studied.
Supersymmetry ensures that the saxions and axinos have the same pattern of the coupling hierarchy as
the clockwork axions. If we assume supersymmetry breaking is universal over the clockwork sites,
the coupling structure is preserved, while the mass orderings of the saxions and axinos can differ depending on the
supersymmetry breaking scale. 
While the massive saxions and axions quickly decay, the lightest axino can be stable and thus a dark matter candidate.
The relic abundance of the axino dark matter from thermal production 
is mostly determined by decays of the heavier axinos in the normal mass ordering. 
This exponentially enhances the thermal yield compared to the conventional axino scenarios. 
Some cosmological issues are discussed.
}
\end{abstract}


\maketitle

\section{Introduction}


One of the strongest beliefs in particle physics is that 
there exist extended sectors of new physics beyond the standard model (SM).
In theoretical aspects, it is invoked to resolve fine-tuning problems residing in the SM.
In practical aspects, the SM does not contain physics for essential phenomena such as neutrino oscillation,
matter-antimatter asymmetry, 
and dark matter (DM).
A widely accepted notion of extensions of the SM is to introduce `dark' sectors which communicate with the SM via feeble interactions leading to
rational explanations to those phenomena.

A prominent fine-tuning problem in the SM is the strong CP problem.
It can be solved by introducing a spontaneously broken Peccei-Quinn symmetry~\cite{Peccei:1977hh}
which involves the QCD axion~\cite{Weinberg:1977ma,Wilczek:1977pj}.
The axion couples to the gluon field strength and dynamically relaxes the QCD $\theta$-term to zero.
Astrophysical observations constrain axion-gauge boson couplings 
(including the axion-gluon coupling)~\cite{Ayala:2014pea,Viaux:2013lha,Fischer:2016cyd,Hamaguchi:2018oqw}
so that the axion couplings are required to be suppressed by an intermediate scale dynamics.
While such large scale can be induced by exotic heavy quarks~\cite{Kim:1979if,Shifman:1979if} 
or tiny coupling with Higgs doublets~\cite{Dine:1981rt,Zhitnitsky:1980tq},
the origin of the hierarchical structure of new physics still remains unanswered.

The clockwork theory presents a plausible mechanism to build hierarchical mass spectra and interactions from 
a series of multiple non-hierarchical ones. 
An early form of the clockwork structure was studied to achieve a trans-Planckian field excursion 
from two sub-Planckian fields in a natural inflation~\cite{Kim:2004rp}. 
In further studies, it was shown that a number of axions with similar decay constants can produce 
an {\it exponentially} large effective scale~\cite{Choi:2014rja,Choi:2015fiu,Kaplan:2015fuy}.
It has been argued that the same mechanism is applicable for more general systems 
with various spins, scales and couplings~\cite{Giudice:2016yja}.
In particular, the clockwork mechanism is able to construct an intermediate scale ($\gtrsim10^9$~GeV) axion decay constant 
from dynamics near the electroweak scale \cite{Higaki:2015jag}.

In the case of the clockwork axion, a global $U(1)^{N+1}$ symmetry spontaneously breaks at scale $f$ and consequently results in $(N+1)$ Goldstone bosons. The global symmetry is explicitly but softly broken by $N$ mass terms with clockwork structure.
This specific structure leaves unbroken $U(1)$ and a corresponding massless degree of freedom.
If the SM sector couples to one end of $(N+1)$ axions (clockwork gears), 
interactions of the massless mode are exponentially suppressed compared to those from the tangible symmetry breaking scale $f$.
Therefore, one can identify the massless degree with the QCD axion and it provides a neat explanation why the axion decay constant is much larger than the electroweak scale.
In this case, the massless degree becomes a good candidate of dark matter as the usual QCD axion while the massive degrees
 quickly decay into visible particles in that they have non-suppressed couplings with the visible sector.
 
Intriguing phenomena in the dark sector (here axion sector) arise if one considers a supersymmetric model of the clockwork axion.
Supersymmetry (SUSY) itself is also an elegant solution to the gauge hierarchy problem which is another fine-tuning problem in the SM.
All pseudo-Nambu-Goldstone bosons (pNGBs)\footnote{The zero mode also becomes a pseudo-Nambu-Goldstone boson once one introduces the interaction with the QCD.} corresponding to $U(1)^{N+1}$ accompany their fermion partners, which we call axinos in this context.
The supersymmetry dictates the same clockwork pattern to axinos and leads to clockwork fermions.
There are more interesting phenomena in the clockwork axinos. 
The $R$-parity, if it is preserved, prevents the heavy axinos from decaying into only the SM particles.
For example, if all the SUSY partners in the SM sector are heavy and only the axinos are $R$-parity odd particles near or below the electroweak scale,
axinos can decay only into another axinos with axions. It leads {\it inter-dark-sector} transitions,
which make all the axino states produced from thermal bath contribute to dark matter number density.

In this paper, we consider a simple model of the supersymmetric clockwork axion,
which consists of $(N+1)$ chiral superfields containing axions, axinos and also saxions (scalar partners of axions).
In the SUSY preserving limit, all three components have the same clockwork structure
for masses and couplings.
Once the SUSY is broken, all three components receive SUSY breaking masses 
and thus masses of saxions and axinos deviate from the axion masses,
while the couplings remain the same clockwork structure.
In a mass spectrum in which the axinos are much lighter than the saxions and axions (except the zero mode axion),
the axinos are domaninatly produced via the gluon scattering mediated by gluinos.
The heavy axinos eventually decay into the lightest axino which is the dark matter in this model.
Furthermore, due to the clockwork structure, 
the axino DM number density is determined by much more enhanced strengths than 
its actual interactions with the SM sector but is {\it independent} of details of the clockwork gears (clockwork parameter and number of gears).


This paper is organized as follows.
In Sec.~\ref{sec:rev_cwax}, we briefly review a clockwork axion model
to show essential elements of the theory.
In Sec.~\ref{sec:susy_ext}, we consider a SUSY extension and 
the mass spectrum for axions, saxions and axinos.
In Sec.~\ref{sec:ther_prod}, we present a complete list of processes for axino production
and the axino abundance in a simple spectrum.
In Sec.~\ref{sec:cos_iss}, we discuss some cosmological issues related to the model.
In Sec.~\ref{sec:conc}, we conclude this paper.





\section{review of clockwork axion}
\label{sec:rev_cwax}

In this section, we briefly review a clockwork axion model to elucidate
essential features of the clockwork theory. 
In the next section, we will supersymmetrize the clockwork axion and see what appears in the model.
We follow a simple formulation shown in Refs.~\cite{Kaplan:2015fuy, Giudice:2016yja}, but 
the basic structure is the same as another formulations in Refs.~\cite{Choi:2014rja, Choi:2015fiu,Higaki:2015jag}.

Let us consider $N+1$ pNGBs originating from a broken global $U(1)^{N+1}$ symmetry.
Below the energy scale $f$ where all $N+1$ $U(1)$ symmetries are broken, Goldstone fields are expressed by
\begin{equation}
U_j=fe^{i\phi_j/(\sqrt{2}f)}.
\end{equation}
The Lagrangian is given by
\begin{eqnarray}
{\cal L}=&&f^2\sum_{j=0}^N\partial_{\mu}U_j\partial^{\mu}U_j
+m^2f^2\sum_{j=0}^{N-1}\left(U_j^{\dagger}U_{j+1}^q+\text{h.c.}\right)+\cdots\nonumber\\
=&&\frac12\sum_{j=0}^{N}\partial_{\mu}\phi_j\partial^{\mu}\phi_j - V(\phi_j),
\label{eq:pngb_lag}
\end{eqnarray}
where the ellipsis denotes higher order terms. 
The potential of $\phi$ fields are given up to the quadratic order by
\begin{eqnarray}
V(\phi_j)=&&-m^2f^2\sum_{j=0}^{N-1}e^{-i(\phi_j-q\phi_{j+1})/\sqrt{2}f}+\text{h.c.}\nonumber\\
=&&\frac12 m^2 \sum_{j=0}^{N-1}(\phi_j-q\phi_{j+1})^2+\cdots,\nonumber\\
=&&\frac12 m^2\sum_{i,j=0}^{N}{\mathbf M}_{{\rm CW} ij}\phi_i\phi_j+\cdots,
\end{eqnarray}
where a matrix ${\mathbf M}_{\rm CW}$ which we call here the clockwork matrix is given by
\begin{equation}
{\mathbf M}_{\rm CW}=
\begin{pmatrix}
1 & -q & 0 & \cdots & & 0 \\
-q & 1+q^2 & -q & \cdots & & 0 \\
0 & -q & 1+q^2 & \cdots & & 0 \\
\vdots & \vdots& \vdots& \ddots & & \vdots \\
& & & & 1+q^2 & -q \\
0 & 0 & 0 & \cdots & -q & q^2 
\end{pmatrix}.
\end{equation}
The matrix is real and symmetric, and thus is diagonalized by an orthogonal matrix $\mathbf O$.
Hence the mass eigenstates $a_j$ satisfies the relation
\begin{equation}
\phi_j={\mathbf O}_{jk}a_k
\end{equation}
with mass eigenvalues given by
\begin{equation}
{\mathbf O}^T {\mathbf M}_{\rm CW} {\mathbf O}= {\rm diag}(\lambda_0,\cdots,\lambda_k).
\label{eq:cw_eig_val}
\end{equation}
The eigenvalues and mixing matrix components are given by
\begin{eqnarray}
&\lambda_0=0, &\lambda_k =q^2+1-2q\cos\left(\frac{k\pi}{N+1}\right),\\
&{\mathbf O}_{j0}=\frac{{\cal N}_0}{q^j}, &{\mathbf O}_{jk}={\cal N}_k\left[q\sin\frac{jk\pi}{N+1}-\sin\frac{(j+1)k\pi}{N+1}\right],
\label{eq:mix_mat}\\
&&\text{for}\quad j=0,\cdots,N;~~ k=1,\cdots,N \ ,\nonumber
\end{eqnarray}
where
\begin{eqnarray}
{\cal N}_0=\sqrt{\frac{q^2-1}{q^2-q^{-2N}}},\quad
{\cal N}_k=\sqrt{\frac{2}{(N+1)\lambda_k}}.
\end{eqnarray}
The axion masses are thus given by $m_{a_j}^2=m^2\lambda_j$.
One can see that one degree remains massless and it corresponds to the $U(1)$
not broken by mass terms in Eq.~\eqref{eq:pngb_lag}.

Suppose that the $N$-th field couples to the SM sector via topological terms, {\it i.e.},
\begin{equation}
{\cal L}=\left[\frac{g_s^2}{32\pi^2}G^b_{\mu\nu}\tilde{G}^{b\mu\nu}+
\frac{g_1^2C_{aYY}}{16\pi^2}B_{\mu\nu}\tilde{B}^{\mu\nu}\right]\frac{\phi_N}{f},
\end{equation}
where $g_s$ and $g_1$ are $SU(3)_c$ and $U(1)_Y$ gauge coupling constants, 
$G^b_{\mu\nu}$, $B_{\mu\nu}$, $\tilde{G}^b_{\mu\nu}$ and $\tilde{B}_{\mu\nu}$  are corresponding gauge field strengths
and their duals, respectively,
and $C_{aYY}$ is a model-dependent constant of order unity.
After clockworking, the above terms lead to interactions between all axions and the SM gauge bosons:
\begin{eqnarray}
{\cal L}=&&\left[\frac{g_s^2}{32\pi^2}G^b_{\mu\nu}\tilde{G}^{b\mu\nu}
+\frac{g_1^2C_{aYY}}{16\pi^2}B_{\mu\nu}\tilde{B}^{\mu\nu}\right]\nonumber\\
&&\times\frac{1}{f}\left(\frac{{\cal N}_0}{q^N} a_0-\sum_{k=1}^{N}(-1)^k{\cal N}_kq\sin\frac{k\pi}{N+1} a_k\right).
\label{eq:axion_coupling}
\end{eqnarray}
One can easily see that the coupling of the zero mode axion is exponentially suppressed compared to that from the actual symmetry breaking scale $f$
while the others are scaled by only $1/N^{3/2}$ for large $N$.
For $q=2$ and $N=20$, the exponential factor is around $10^6$, so one can achieve a good QCD axion
even from $f=1$~TeV.

If the zero mode is the QCD axion, it finally becomes massive by the chiral symmetry breaking in the strong sector of the SM, but the mass is still tiny.
As is well known, the QCD axion has very long lifetime, so it could be a dark matter component.
On the other hand, massive states are rather strongly coupled to the SM sector. 
One can obtain decay widths of the massive modes to the photon pair as
\begin{eqnarray}
\Gamma_{a_k\to \gamma\gamma}=&&\frac{C_{a\gamma\gamma}^2\alpha_{\rm em}^2}{256\pi^3}
{\cal N}_k^2q^2\sin^2\frac{k\pi}{N+1}
\frac{m_{a_k}^3}{f^2}\nonumber\\
\sim&&(10^{-7}~\text{s})^{-1}\left(\frac{20}{N}\right)^3\left(\frac{10~\text{TeV}}{f}\right)^2
\left(\frac{m}{\text{GeV}}\right)^3 \nonumber\\
\end{eqnarray}
where $\alpha_{\rm em}$ is the fine structure constant and $C_{a\gamma\gamma}$ is a constant determined by $C_{aYY}$ and chiral symmetry breaking effect ({\it e.g.}, $C_{a\gamma\gamma}\simeq-1.92$ for Kim-Shifman-Vainshtein-Zakharov (KSVZ) model~\cite{diCortona:2015ldu}).
These states decay before the big bang nucleosynthesis (BBN) for $f=10$~TeV and $m=1$~GeV.
In most cases, therefore, the massive states do not make significant impacts on the evolution of the universe.

\section{A supersymmetric extension}
\label{sec:susy_ext}

In this section, we consider a SUSY extension of the clockwork axion model.

\subsection{A model}

Similar to a simple construction in Ref.~\cite{Kaplan:2015fuy},
one can consider a K\"ahler potential and a superpotential 
\begin{eqnarray}
K&=&\sum_{j=0}^N \left( X_j^{\dagger}X_j+Y_j^{\dagger}Y_j+Z_j^{\dagger}Z_j \right), \label{kahlerpot}\\
W&=&\sum_{j=0}^N\kappa Z_j\left(X_jY_j-v^2\right)\nonumber\\
&&+\frac{1}{v^{q-1}}\sum_{j=0}^{N-1}\left(m X_jY_{j+1}^q+m' Y_jX_{j+1}^q\right), \label{superpot}
\end{eqnarray}
where charge assignment of $Z_j$, $X_j$, and $Y_j$ under $U(1)_j$ is $(0, +1, -1)$.
The first term reflects the spontaneous breaking of $U(1)$ global symmetry
near $v$ while the second term corresponds to a small explicit breaking effect for $m, m'\ll v$.
We consider a generic case for $m\ne m'$ leading to $\langle X_j\rangle\ne \langle Y_j\rangle$
which is important for inter-dark-sector couplings in Eq.~\eqref{eq:inter_trans}.
The fields are stabilized at
\begin{equation}
\langle Z_j\rangle=-\frac{q+1}{\kappa}\sqrt{mm'},\quad \langle X_j\rangle =x,\quad \langle Y_j\rangle=y
\end{equation}
where\footnote{Here we can take a field basis where all parameters are taken to be real and positive
except $\kappa$. In this basis, the supersymmetric effective action for the axion supermultiplets does not involve any complex parameter as we will see below.}  
\begin{equation}
xy=v^2,\quad x=\left(\frac{m}{m'}\right)^{\frac{1}{2(q-1)}}v.
\end{equation} 
Below the spontaneous $U(1)$ symmetry breaking scale, this theory can be described by
chiral superfields containing pNGBs,
\begin{equation}
\Phi_j=\frac{1}{\sqrt{2}}(\sigma_j+i \phi_j)+\sqrt{2}\theta \psi_j + \theta^2 F_j,
\end{equation}
where $\sigma_j$ and $\psi_j$ are scalar and fermion partners of $\phi_j$.
One can write
\begin{equation}
X_j=x ~e^{\Phi_j/v_0},\quad Y_j=y ~e^{-\Phi_j/v_0},
\end{equation}
where $v_0=\sqrt{x^2+y^2}$.
The effective K\"ahler potential and superpotential become
\begin{eqnarray}
K_{\rm eff}&=&v_0^2
\sum_{j=0}^{N}\left[
\cosh\left(\frac{\Phi_j+\Phi_j^{\dagger}}{v_0}\right)\right.\nonumber\\
&&\left.+\xi\sinh\left(\frac{\Phi_j+\Phi_j^{\dagger}}{v_0}\right)
\right],
\label{eq:kahl_eff}\\
W_{\rm eff}&=& m_\Phi v_0^2\sum_{j=0}^{N-1}\cosh\left(\frac{\Phi_j-q\Phi_{j+1}}{v_0}\right),
\label{eq:sup_eff}
\end{eqnarray}
where $\xi=(x^2-y^2)/v_0^2$ and
\dis{
m_\Phi \equiv 2\sqrt{mm'} \left(\frac{v}{v_0}\right)^2.
}
In the K\"ahler potential, we have omitted $Z^{\dagger} Z$ since it is irrelevant in the low energy dynamics.
The above superpotential shows that
the supersymmetric minimum is achieved for $\langle \Phi_j - q\Phi_{j+1}\rangle=0$
and the supersymmetric mass term indeed has the clockwork structure
proportional to an overall mass scale $m_{\Phi}$.
One can obtain superfields in the eigenbasis with mixing matrix in Eq.~\eqref{eq:mix_mat}:
\begin{equation}
\Phi_i={\mathbf O}_{ij}A_j.
\end{equation}
Hence one {\it supermultiplet} remains massless after clockworking.

Similarly to the clockwork axion model, one can introduce couplings of the $N$-th superfield to the SM gauge fields as
\begin{eqnarray}
{\cal L}=&&-\frac{g_s^2}{32\pi^2}\frac{C_{aGG}}{v_0}\int d^2\theta \Phi_N {\cal W}^{b\alpha}{\cal W}^b_{\alpha}+\text{h.c.}\nonumber\\
&&-\frac{g_1^2}{16\pi^2}\frac{C_{aYY}}{v_0}\int d^2\theta \Phi_N {\cal W}^{\alpha}{\cal W}_{\alpha}+\text{h.c.},
\label{eq:susy_ax_coupling}
\end{eqnarray}
where ${\cal W}^b$ is the gluon superfield, ${\cal W}$ is the hypercharge superfield, and $C_{aGG}$ and $C_{aYY}$ are
model-dependent coefficients of the order of unity.
After clockworking, the zero mode superfield has exponentially suppressed interactions as
\begin{eqnarray}
{\cal L}=&&-\frac{g_s^2}{32\pi^2}\frac{C_{aGG}}{f_0}\int d^2\theta A_0 {\cal W}^{b\alpha}{\cal W}^b_{\alpha}+\text{h.c.}\nonumber\\
&&-\frac{g_1^2}{16\pi^2}\frac{C_{aYY}}{f_0}\int d^2\theta A_0 {\cal W}^{\alpha}{\cal W}_{\alpha}+\text{h.c.},
\end{eqnarray}
where $f_0=q^Nv_0$.

\subsection{SUSY breaking effects and mass spectrum}

Once the SUSY is broken, the mass spectrum for each component alters.
The pNGBs and scalar partners would receive mass contributions from 
SUSY breaking in the superpotential as
\begin{eqnarray}
{\cal L}&=&\int d \theta^2 (1+m_s\theta^2) W+\text{h.c.}\nonumber\\
 \rightarrow V&=&-m_\Phi |m_s| v_0^2 \nonumber\\
&&\times\sum_{j=0}^{N-1}\left[ e^{(\sigma_j-q\sigma_{j+1})/\sqrt{2}v_0} \cos\left(\frac{\phi_j-q\phi_{j+1}}{\sqrt{2}v_0} +\delta_s \right) \right.\nonumber\\
&&\left. +e^{-(\sigma_j-q\sigma_{j+1})/\sqrt{2}v_0} \cos\left(\frac{\phi_j-q\phi_{j+1}}{\sqrt{2}v_0} -\delta_s \right) \right], \nonumber \\
\label{eq:sb_eff}
\end{eqnarray}
where $\delta_s$ is the complex phase of $m_s$. 
For simplicity, we will focus on parameter space where
vacuum field configuration is close to the supersymmetric minimum point $\langle \Phi_j - q\Phi_{j+1}\rangle=0$.
Near the point,
the above potential becomes approximately
\begin{eqnarray}
V_{\sigma}&\simeq&-2m_\Phi |m_s| v_0^2\cos\delta_s\sum_{j=0}^{N-1}\cosh\left(\frac{\sigma_j-q\sigma_{j+1}}{\sqrt{2}{v_0}}\right),\\
V_{\phi}&\simeq&-2m_\Phi |m_s| v_0^2\cos\delta_s\sum_{j=0}^{N-1}\cos\left(\frac{\phi_j-q\phi_{j+1}}{\sqrt{2}v_0}\right)
\end{eqnarray}
along the scalar and pNGB directions, respectively.
It contributes to {\it squared masses} with the clockwork structure for the pNGBs and their scalar partners.
The mass scale for this contribution is determined by
\dis{
m_{\rm sb}^2 \equiv m_\Phi |m_s|\cos\delta_s. 
}
If SUSY breaking effects also arise in the K\"ahler potential in Eq.~\eqref{eq:kahl_eff},
scalars and fermions acquire additional masses
which are {\it diagonal} in the basis of chiral superfields.
We write $m_{\sigma}^K$ and $m_{\psi}^K$, repectively, for the scalars and fermions.
We further assume these terms are the same for all $j$'s,
and thus the mass matrices from this contribution are proportional to the identity matrix. 
While it is expected to have  $m_{\sigma}^K\sim m_{\psi}^K$ in generic cases, it is possible to have $m_{\sigma}^K\gg m_{\psi}^K$
in some cases.\footnote{We refer readers to Ref.~\cite{Goto:1991gq,Chun:1992zk,Chun:1995hc,Bae:2014efa} for general discussion for the mass generation and Ref.~\cite{Abe:2001cg,Nakamura:2008ey} for explicit models with $m_{\sigma}^K\gg m_{\psi}^K$.}

Mass spectra for the pNGBs, scalars and fermions are summarized as
\begin{eqnarray}
{\mathbf M}^2_{\phi}&=&m_{\Phi}^2{\mathbf M}^2_{\rm CW}
+m_{\rm sb}^2{\mathbf M}_{\rm CW}, \label{Mat_a}\\
{\mathbf M}^2_{\sigma}&=&m_{\Phi}^2{\mathbf M}^2_{\rm CW}
-m_{\rm sb}^2{\mathbf M}_{\rm CW}+\left(m_{\sigma}^{K}\right)^2{\mathbf I}, \label{Mat_s}\\
{\mathbf M}_{\psi}&=&m_{\Phi}{\mathbf M}_{\rm CW}+m_{\psi}^{K}{\mathbf I}. \label{Mat_an}
\end{eqnarray}
The $(N+1)\times(N+1)$ identity matrix is denoted by $\mathbf I$.
We emphasize that all the mass matrices are diagonalized by the same mixing matrix in Eq.~\eqref{eq:mix_mat}.
Hence 
we write mass eigenstates
\begin{eqnarray}
\phi_j&=&{\mathbf O}_{jk}a_k, \label{cw_a}\\
\sigma_j&=&{\mathbf O}_{jk}s_k, \label{cw_s}\\
\psi_j&=&{\mathbf O}_{jk}\tilde{a}_k, \label{cw_an}
\end{eqnarray}
with mass eigenvalues
\begin{eqnarray}
m_{a_k}^2&=&m_{\Phi}^2\lambda_k^2+m_{\rm sb}^2\lambda_k, \label{mass_a}\\
m_{s_k}^2&=&m_{\Phi}^2\lambda_k^2-m_{\rm sb}^2\lambda_k+(m_{\sigma}^K)^2, \label{mass_s}\\
m_{\tilde{a}_k}&=&m_{\Phi}\lambda_k+m_{\psi}^K \label{mass_an}
\end{eqnarray}
and call these states axions, saxions and axinos, respectively.
While the zero mode axion, $a_0$ is massless in that the mass term is determined only by $\lambda_0$,
both $s_0$ and $\tilde{a}_0$ become massive due to the SUSY breaking effect in the K\"ahler potential.
While $m_{\Phi}^2$ is always positive by definition, $m_{\rm sb}^2>-m_{\Phi}^2(q-1)^2$  is required not to destabilize axion directions.
Once this condition is satisfied, 
the mass difference $\delta m_{a_k}^2\simeq m_{a_{k+1}}^2- m_{a_k}^2$ is given by
\begin{eqnarray}
\delta m_{a_k}^2&>&2q m_{\Phi}^2\left[\lambda_{k+1}\left(1-\cos\frac{(k+1)\pi}{N+1}\right)\right.\nonumber\\
&&\left.-\lambda_{k}\left(1-\cos\frac{k\pi}{N+1}\right)\right].
\end{eqnarray}
Since $\lambda_{k+1}>\lambda_k$ 
and the cosine is monotonically decreasing, $\delta m_{a_k}^2$ is always positive.
Thus the ordering of axion mass eigenvalues is the same as that in Eq.~\eqref{eq:cw_eig_val}, although
mass differences alter.
On the other hand, the ordering of eigenvalues can be different for the saxions and axinos.
If $m_{\rm sb}^2\gg m_{\Phi}^2$ ({\it i.e.} $|m_s| \cos \delta_s \gg m_\Phi$), the $\lambda_{k}$-dependent part becomes negative
so as to destabilize the supersymmetric vacuum.
Yet if
$(m_{\sigma}^K)^2$ is large enough, the supersymmetric vacuum can be maintained.
In this case, the largest eigenvalue is $m_{s_0}^2$ while the smallest one is $m_{s_N}^2$.
The mass ordering of the saxions is {\it inverted} when being compared to that of the axions.
The same thing happens for the axinos.
If $m_{\psi}^K<0$, $\tilde{a}_0$ may not be the lightest mode.
In the case $|m_{\psi}^K| > m_{\Phi} \lambda_N$ with negative $m_{\psi}^K$, the mass ordering of the axinos 
is inverted.
The ordering may be even not monotonic if $|m_{\psi}^K|<m_{\Phi}\lambda_N$.
Nevertheless, we consider the `normal' hierarchy, {\it i.e.}, 
$m_{s_0}^2<\cdots<m_{s_N}^2$ and $m_{\tilde{a}_0}<\cdots<m_{\tilde{a}_N}$ in later discussion.

Some comments are in order about conditions to get the clockwork mixing pattern in Eqs. (\ref{cw_a})-(\ref{cw_an}), which is crucial
for exponential coupling hierarchy. 
In the limit of $m,m'\to 0$, the global $U(1)^{N+1}$ symmetry is preserved and thus there exist $N+1$ chiral superfields, $\Phi_j$,
corresponding to $N+1$ flat directions, $X_jY_j=v^2$.
Once $m$ and $m'$ are turned on, the global $U(1)^{N+1}$ symmetry is broken down to $U(1)$.
The remaining $U(1)$ symmetry leaves one flat direction while the others become massive.
It can be explicitly seen by the fact that
the superpotential does not change under
\dis{
\Phi_j \rightarrow \Phi_j + q^{-j} \alpha
}
with a constant $\alpha$. 
This ensures the superfield corresponding to the remaining flat direction to have exponentially small couplings. 
The SUSY breaking in the superpotential (\ref{eq:sb_eff}) also respects it, so the flat direction remains.
On the other hand, the SUSY breaking in the K\"ahler potential 
develops masses of the scalars and fermions, while the masses do not respect the above symmetry.
This means that except the axion, the saxion and axino may not get small couplings 
if the SUSY breaking effect in the K\"ahler potential is significant. 
More quantitatively those SUSY breaking contributions for their mass matrices $(m_\sigma^K)_{ij}$ and $(m_\psi^K)_{ij}$ have to
be sufficiently small compared to $m_\Phi$ or $m_{\rm sb}$, or closely proportional to the identity matrix as in Eqs. (\ref{Mat_s}) and (\ref{Mat_an})
in order to preserve the clockwork coupling hierarchy. The hierarchy would be spoiled if departure from being proportional to the identity matrix
is of the order of $m_\Phi$ or $m_{\rm sb}$.
This argument is valid even when the supersymmetric parameters $\kappa, v, m, m'$
in (\ref{superpot}) and the SUSY breaking parameter $m_s$ in (\ref{eq:sb_eff}) are dependent on sites $j$. 
Such dependency makes a difference only on mass eigenvalues in Eqs. (\ref{mass_a}) - (\ref{mass_an}) without qualitatively changing
our results.

Let us finally make a remark for a benchmark spectrum. 
If we want to identify the zero mode axion $a_0$ as QCD axion
with an intermediate scale decay constant,
$v_0$ can be as low as $O(1)$ TeV for $N\lesssim 20$.
Effective descriptions in Eqs.~\eqref{eq:kahl_eff}, \eqref{eq:sup_eff}, and \eqref{eq:sb_eff} 
are valid only for $m,m',m_s \ll v_0$.
Hence all states are expected to be near or below the weak scale.

\subsection{Interactions}

The axions have the same interactions as in the case of the non-SUSY model in Eq.~\eqref{eq:axion_coupling}.
The saxions also have similar interactions from the SUSY coupling term in Eq.~\eqref{eq:susy_ax_coupling}.
The saxion-gauge boson interactions are given by
\begin{eqnarray}
{\cal L}_{\rm sax}=&&\left[\frac{g_s^2 C_{aGG}}{32\pi^2}G^b_{\mu\nu}G^{b\mu\nu}
+\frac{g_1^2C_{aYY}}{16\pi^2}B_{\mu\nu}B^{\mu\nu}\right]\nonumber\\
\times\frac{1}{\sqrt{2}v_0}&&\left(\frac{{\cal N}_0}{q^N} s_0-\sum_{k=1}^{N}(-1)^k{\cal N}_kq\sin\frac{k\pi}{N+1} s_k\right).
\label{eq:sax_ax_gauge}
\end{eqnarray}
We neglect axion-gluino, saxion-gluino and saxion-squark interactions derived from Eq.~\eqref{eq:susy_ax_coupling}
since they are irrelevant in the later discussion.
The axino interactions are derived in the same way:
\begin{eqnarray}
{\cal L}_{\rm axn}&&=\frac{1}{\sqrt{2} v_0}
\left(
\frac{{\cal N}_0}{q^N} \bar{\tilde{a}}_0-\sum_{k=1}^{N}(-1)^k{\cal N}_kq\sin\frac{k\pi}{N+1} \bar{\tilde{a}}_k
\right)\nonumber\\
&&\times
\left(\frac{g_s^2 C_{aGG}}{32\pi^2}G_{\mu\nu}^b\sigma^{\mu\nu}\gamma^5\tilde{g}^b
+\frac{g_1^2 C_{aYY}}{16\pi^2}B_{\mu\nu}\sigma^{\mu\nu}\gamma^5\tilde{B}
\right), \nonumber \\
\label{eq:axn-gau_int}
\end{eqnarray}
where $\sigma^{\mu\nu}\equiv\frac{i}{2}[\gamma^{\mu},\gamma^{\nu}]$.
The gluino and bino are denoted by $\tilde{g}$ and $\tilde{B}$.
It is noteworthy that we use Majorana spinors for axinos and gauginos in Eq.~\eqref{eq:axn-gau_int} and the later discussion.

In addition, the K\"ahler potential in Eq.~\eqref{eq:kahl_eff} generates qubic (and also higher-order) interactions 
between the axions, saxions and axinos:
\begin{eqnarray}
K&\supset& \frac{\xi}{3!}v_0^2\sum_{j=0}^N\left(\frac{\Phi_j+\Phi_j^{\dagger}}{v_0}\right)^3\nonumber\\
\rightarrow {\cal L}_{nml}&=&\frac{\xi}{\sqrt{2}v_0}\sum_{j}^N {\mathbf O}_{jn}{\mathbf O}_{jm}{\mathbf O}_{jl}\nonumber\\
&&\times\left[s_n(\partial_{\mu} a_m)(\partial^{\mu}a_l)+s_n(\partial_{\mu}s_m)(\partial^{\mu}s_l)\right.\nonumber\\
&&\left.+is_n\bar{\tilde{a}}_m\gamma^{\mu}\partial_{\mu}\tilde{a}_l
-(\partial_{\mu}a_n)\bar{\tilde{a}}_m\gamma^5\gamma^{\mu}\tilde{a}_l\right].
\label{eq:inter_trans}
\end{eqnarray}
From this Lagrangian, one can easily read off all trilinear interactions
which mediates {\it inter-dark-sector} transitions.
Here we assume $F_j=0$ for all $j$'s.

\section{thermal production of axinos}
\label{sec:ther_prod}

In this section, we discuss thermal production of axinos in the early Universe.
Since the whole dark sector ({\it i.e.}, axion supermultiplets) 
communicates with the SM sector via the interactions in Eq.~\eqref{eq:susy_ax_coupling} and clockworking,
all the axions, saxions and axinos are produced from thermal plasma after the primordial inflation.
In a SUSY extension, the axinos are odd while the saxions and axions are even
under the $R$-parity if it is preserved.
Therefore the lightest axino can be a dark matter candidate
if it is the lightest $R$-parity odd particle.
The saxions and axions except $a_0$, however, would normally disappear by decaying into 
another light species such as gluons and photons.
In this respect, axino production is more prominent than the others 
for dark matter physics.
We focus on how axinos are produced.

The axino production consists of the following channels: 1) gluino-mediated process,
2) saxion/axion-mediated process, and 3) production from saxion/axino decay.
In particular, we will consider a relatively low reheat temperature $T_R$ below the SUSY breaking scale
so that axino production is mainly from the SM thermal bath. The reason is that the thermal yield of the lightest axino
can easily saturate the DM abundance enhanced by a certain power of the clockwork factor $q^N$ compared to the conventional scenarios
as we will see.

\subsection{Gluino-mediated process}

From the interactions with gauge bosons in Eq.~\eqref{eq:axn-gau_int},
axinos can be produced from the thermal plasma.
If the temperature is larger than masses of the SUSY particles in the SM sector,
the single-axino production is the dominant process which includes the other SUSY particles in either
the initial or final state. 
This scenario has been intensively studied both for the KSVZ-type model~\cite{Covi:1999ty,Covi:2001nw,Brandenburg:2004du,Strumia:2010aa} and for the Dine-Fischler-Srednicki-Zhitnitsky (DFSZ)-type model~\cite{Chun:2011zd,Bae:2011jb,Bae:2011iw}.
If the temperature is smaller than masses of the SUSY particles in the SM sector but still larger than the axino mass, 
{\it e.g.}, $m_{\tilde{a}}\ll T\ll m_{\tilde{g}}\sim m_{\tilde{q}}$,
the single-axino production is Boltzmann-suppressed. 
Instead, the axino pair production becomes more important~\cite{Choi:2018lxt}.
By integrating out the gluino field in Eq.~\eqref{eq:axn-gau_int},
one can obtain an effective Lagrangian for the axino pair production, {\it i.e.}, $gg\to\tilde{a}_n\tilde{a}_m$:
\begin{eqnarray}
{\cal L}_{gg\tilde{a}\tilde{a}}=&&
-\frac{\alpha_s^2 C_{aGG}^2}{1024\pi^2v_0^2m_{\tilde{g}}}
{\mathbf O}_{Nn}{\mathbf O}_{Nm}\nonumber\\
&&\times \bar{\tilde{a}}_n[\gamma^{\mu},\gamma^{\nu}][\gamma^{\rho},\gamma^{\sigma}]\tilde{a}_m
G_{\mu\nu}^bG_{\rho\sigma}^b.
\end{eqnarray}
The squared amplitude for this process is given by
\begin{equation}
|{\cal M}^{\tilde{g}}_{nm}|^2=\frac{\alpha_s^4 C_{aGG}^4}{16\pi^4v_{0}^4m_{\tilde{g}}^2}\left|{\mathbf O}_{Nn}{\mathbf O}_{Nm}\right|^2s^3(1+\cos\theta)^2,
\end{equation} 
where $s$ is the square of the center of mass energy and $\theta$ is the angle between the incoming gluon and outgoing axino.
Here we have summed over all possible degrees of freedom for both the initial and final states.

\subsection{Saxion/axion-mediated process}

Another channel for the axino pair production is realized by the saxion- or axion-mediated processes.
The interactions in Eqs.~\eqref{eq:sax_ax_gauge} and \eqref{eq:inter_trans} lead to
a scattering process $gg\to \left(s_l^* ~\text{or}~a_l^*\right) \to \tilde{a}_n\tilde{a}_m$, and
its squared amplitude is given by
\begin{eqnarray}
\left|{\cal M}^{s/a}_{nm}\right|^2
=&&\frac{\xi^2\alpha_s^2 C_{aGG}^2}{2\pi^2v_{0}^4}
\left|\sum_{l,j}{\mathbf O}_{Nl}{\mathbf O}_{jl}{\mathbf O}_{jn}{\mathbf O}_{jm}\left(\frac{1}{s-m_{l}^2}\right)\right|^2\nonumber\\
&&\times (m_{\tilde{a}_n}+m_{\tilde{a}_m})^2s^3,
\end{eqnarray}
where $m_{l}$ is a mass of $s_l$ or $a_l$.
If $s\gg m_{l}^2$,  the squared amplitude is further simplified, so one can find
\begin{eqnarray}
\left|{\cal M}^{s/a}_{nm}\right|^2 \simeq&&
\frac{\xi^2\alpha_s^2 C_{aGG}^2}{2\pi^2v_{0}^4}
\left|{\mathbf O}_{Nn}{\mathbf O}_{Nm}\right|^2\nonumber\\
&&\times(m_{\tilde{a}_n}+m_{\tilde{a}_m})^2s,
\end{eqnarray}
where we have used an identity
\begin{eqnarray}
\sum_{l,j}{\mathbf O}_{Nl}{\mathbf O}_{jl}{\mathbf O}_{jn}{\mathbf O}_{jm}
={\mathbf O}_{Nn}{\mathbf O}_{Nm}.
\end{eqnarray}
If $s\ll m_{l}^2$, the squared amplitude is approximately given by
\begin{eqnarray}
\left|{\cal M}^{s/a}_{nm}\right|^2 \simeq&&
\frac{\xi^2\alpha_s^2 C_{aGG}^2}{2\pi^2v_{0}^4m_{\rm s/a}^4}
\left|{\mathbf O}_{Nn}{\mathbf O}_{Nm}\right|^2\nonumber\\
&&\times(m_{\tilde{a}_n}+m_{\tilde{a}_m})^2s^3
\end{eqnarray}
where we have assumed $m_l\sim m_{s/a}$ for all $l$,
{\it i.e.}, all masses are of the same order.
In this argument, we have also neglected the zero mode axion contribution since 
its coupling is exponentially suppressed.

\subsection{Production from saxion/axion decay}

Because of the interactions in Eq.~\eqref{eq:inter_trans}, saxions and axions can decay into axino pairs.
One can easily find their partial decay widths:
\begin{eqnarray}
\Gamma(s_l/a_l\to\tilde{a}_n\tilde{a}_m)=&&\frac{\xi^2m_{l}}{16\pi v_0^2}(m_{\tilde{a}_n}+m_{\tilde{a}_m})^2\nonumber\\
\times&&\left|\sum_{j}{\mathbf O}_{jl}{\mathbf O}_{jm}{\mathbf O}_{jn}\right|^2\Delta_{nm},
\end{eqnarray}
where $\Delta_{nm}=1$ (1/2) for $n\ne m$ ($n=m$).
Meanwhile, saxions and axions can also decay into gluon pairs with the partial decay widths
\begin{eqnarray}
\Gamma(s_l/a_l\to gg)=\frac{\alpha_s^2 C_{aGG}^2 m_l^3}{64\pi^3 v_0^2}|{\mathbf O}_{Nl}|^2.
\end{eqnarray}
For $m_{\tilde{a}_n}+m_{\tilde{a}_m}\ll m_l$, saxions and axions decay dominantly into gluons.


\subsection{Secluded spectrum}

Comparing the gluino-mediated and saxion/axion-mediated processes,
the relative ratio between squared amplitudes is given by
\begin{eqnarray}
R\equiv\frac{|{\cal M}^{\tilde{g}}_{nm}|^2}{|{\cal M}^{s/a}_{nm}|^2}\sim
\frac{\alpha_s^2 C_{aGG}^2}{8\pi^2\xi^2}\frac{s^2}{m_{\tilde{g}}^2(m_{\tilde{a}_n}+m_{\tilde{a}_m})^2}
\end{eqnarray}
for $s\gg m_{s/a}^2$,
or
\begin{eqnarray}
R\sim
\frac{\alpha_s^2 C_{aGG}^2}{8\pi^2\xi^2}\frac{m_{s/a}^4}{m_{\tilde{g}}^2(m_{\tilde{a}_n}+m_{\tilde{a}_m})^2}
\end{eqnarray}
for $s\ll m_{ s/a}^2$.
Thus, for $m_{s/a}\gg \sqrt{m_{\tilde{g}}(m_{\tilde{a}_n}+m_{\tilde{a}_m})}$, the gluino-mediated process 
dominates over the saxion/axion-mediated process 
if the reheat temperature $T_R$ is smaller than $m_{\tilde{g}}$.
In this respect, we consider a simple particle mass spectrum with $m_{\tilde{a}_n}\ll m_{s/a}\ll m_{\tilde{g}}$
and $m_{s/a}\gg \sqrt{m_{\tilde{g}}(m_{\tilde{a}_n}+m_{\tilde{a}_m})}$.
In this spectrum, moreover, the branching fraction of $s/a\to\tilde{a}_n\tilde{a}_m$
is highly suppressed by small axino masses compared to saxion and axion masses.
Because of the supersymmetry, saxions, axions and axinos are produced with the similar amount
in the large $T_R$ limit,
so the amount of axinos from saxion and axion decays is negligible in this case.
Hence axinos are predominantly produced in pairs via the gluino-dominated process.
We call this a `secluded' spectrum.

\subsection{Thermal yield of axinos}

One can obtain the thermal-averaged axino production cross section from the squared amplitude.
For a $\tilde{a}_n\tilde{a}_m$ pair production, the thermal-averaged cross section is given by
\begin{eqnarray}
\langle \sigma v\rangle_{nm} \simeq \frac{6\alpha_s^4C_{aGG}^4 T^4}{\pi^5[\zeta(3)]^2 v_{0}^4m_{\tilde{g}}^2}
\left|{\mathbf O}_{Nn}{\mathbf O}_{Nm}\right|^2\Delta_{nm},
\end{eqnarray}
where $T$ is the plasma temperature and $\zeta$ is the zeta function.
The yield of $\tilde{a}_n$ state, $Y_{\tilde{a}_n}\equiv n_{\tilde{a}_n}/s$ 
($n_{\tilde{a}_n}$: number density of ${\tilde{a}_n}$, $s$: entropy density)
is then given by
\begin{equation}
Y_{\tilde{a}_n}\simeq\left(\frac{3\sqrt{10}}{[g(T_R)]^{3/2}}\right)
\frac{243\alpha_s^4 C_{aGG}^4 M_PT_R^5}{16\pi^{12}v_{0}^4m_{\tilde{g}}^2}
\left|{\mathbf O}_{Nn}\right|^2,
\end{equation}
where $g(T_R)$ is the effective degrees of freedom at $T_R$
and $M_P$ is the reduced Planck mass.
Here we have used an identity
\begin{equation}
\sum_{m}\left|{\mathbf O}_{Nm}\right|^2=1
\label{eq:iden_orth}
\end{equation}
It is noteworthy that we have included the correction from 
the continuous reheating process~\cite{Garcia:2017tuj}.

In the secluded spectrum, the heavier axinos eventually decay into the lightest axino,
so the final yield of {\it axino dark matter} is determined by the sum of all the axino yields:
\begin{eqnarray}
Y_{\tilde{a}}^{\rm DM}&=&\sum_{n}Y_{\tilde{a}_n}\nonumber\\
&\simeq&
\left(\frac{3\sqrt{10}}{[g(T_R)]^{3/2}}\right)
\frac{243\alpha_s^4C_{aGG}^4M_PT_R^5}{16\pi^{12}v_{0}^4m_{\tilde{g}}^2},
\end{eqnarray}
where we have used the identity in Eq.~\eqref{eq:iden_orth}.
The axino DM abundance is thus given by
\begin{eqnarray}
\Omega_{\tilde{a}}h^2&\simeq& 2.8\times 10^5\times Y^{\rm DM}_{\tilde{a}}
\left(\frac{m_{\tilde{a}}}{\text{MeV}}\right)\nonumber\\
&\simeq&0.13\times \left(\frac{C_{aGG}}{1} \right)^4\left(\frac{\text{TeV}}{v_0}\right)^4
\left(\frac{10~\text{TeV}}{m_{\tilde{g}}}\right)^2\nonumber\\
&&\times\left(\frac{T_R}{40~\text{GeV}}\right)^5
\left(\frac{m_{\tilde{a}}}{10~\text{keV}}\right), \label{DM_yield}
\end{eqnarray}
where we have used $\alpha_s\simeq0.1$ and $m_{\tilde{a}}$ denotes the lightest axino mass.

In the normal hierarchy, $\tilde{a}_0$ is the lightest axino state and thus dark matter.
Its interaction to the SM sector is highly suppressed by $1/q^N$, so most of the DM axinos
are produced via decays of the heavier axinos
which have interactions being mildly scaled by $\sim 1/N^{3/2}$.
In other words,
the clockwork mechanism realizes largely enhanced axino production
in spite of the feebly interacting nature of DM species.
Compared to the conventional non-clockwork scenarios of the same axino coupling to the SM, 
the DM abundance is enhanced by the factor $(f_0/v_0)^4 = q^{4N}$.

\section{Cosmological issues}
\label{sec:cos_iss}

\subsection{Heavy axino decays}

As discussed in Sec.~\ref{sec:ther_prod}, 
most of the DM axinos are produced via decays of the heavier axinos.
In the secluded spectrum, an axino
can decay into a lighter axino plus the zero mode axion, {\it i.e.}, $\tilde{a}_n\to\tilde{a}_m+a_0$, $n>m$
due to the interaction in Eq.~\eqref{eq:inter_trans}.
The decay width is given by
\begin{eqnarray}
\Gamma(\tilde{a}_n\to\tilde{a}_m+a_0)=&&\frac{1}{16\pi}\frac{\xi^2}{v_{0}^2}\left|\sum_{j=0}^N
{\mathbf O}_{j0}{\mathbf O}_{jn}{\mathbf O}_{jm}\right|^2\nonumber\\
&&\times m_{\tilde{a}_n}^3\left(1-\frac{m_{\tilde{a}_m}^2}{m_{\tilde{a}_n}^2}\right)^{3},
\end{eqnarray}
While the DM axino yield is independent of the decay path,
the phase space distribution of the DM axinos is highly dependent
on the decay path, lifetimes and mass differences.
Depending on the model parameters $N$, $q$, $m_{\Phi}$ and $m_{\psi}^K$, 
the resulting phase space distribution can deviate from the conventional thermal distribution.
Hence, it may impact on the structure formation~\cite{bky2020}.

\subsection{Axion string-wall network}

Since the clockwork axions and saxions have short lifetimes,
their cosmological population from initial misalignment quickly decays without leaving  substantial impacts. 
However a network of axion strings and domain walls formed by the global $U(1)^{N+1}$ symmetry breaking
can sizably contribute to the dark radiation \cite{Long:2018nsl} and yield observable gravitational waves \cite{Higaki:2016jjh}. 
In Ref.~\cite{Long:2018nsl} it is argued that the axion DM production from collapse of the string-wall network of the clockwork gears is negligible
due to the suppressed interactions between the axion and clockwork gears. Yet relativistic axions produced from the clockwork gear domain wall
contribute to dark radiation at the recombination epoch as
\dis{
\Delta N_{\rm eff} \simeq &0.1 \left( \frac{v_\omega}{1}\right)^2 
\left( \frac{m_{\Phi}}{10 \,\textrm{TeV}}\right) \left(\frac{v_0}{10^6 \,\textrm{GeV}} \right)^2 \\
&\times \left(\frac{g_{*S}(T_{a_0})}{20} \right)^{-4/3} \left(\frac{T_{a_0}}{0.2\, {\rm GeV}}\right)^{-2} 
} 
where $T_{a_0}$ is the temperature
at which the axion $a_0$ gets a mass, and $v_\omega \leq 1$ parametrizes the spectrum of small-scale perturbations on the domain wall. For $T_{a_0}$, we use the value of QCD axion as the normalization.  
Observations of the comic microwave background require $\Delta N_{\rm eff} \lesssim 0.1$ \cite{Ade:2015xua}.
Thus it sets an upper bound on the quantity $m_{\Phi} v_0^2$ for a given $T_{a_0}$. In fact, this quantity corresponds to the domain wall tension. 
On the other hand, the violent annihilation of the clockwork domain walls gives rise to gravitational waves of frequencies of the order of the Hubble parameter. 
It turns out that 
we have a similar observational constraint on the domain wall tension~\cite{Higaki:2016jjh}.
Using the estimation of \cite{Higaki:2016jjh}, to be consistent with pulsar timing observations \cite{Lentati:2015qwp, Arzoumanian:2018saf, Lasky:2015lej, Kerr:2020qdo, Perera:2019sca}, 
our model parameters need to satisfy
\begin{eqnarray}
&& \left( \frac{m_{\Phi}}{10 \,\textrm{TeV}}\right) \left(\frac{v_0}{10^6 \,\textrm{GeV}} \right)^2  \nonumber\\
&\lesssim& 0.1 \left( \frac{\epsilon_{\rm gw}}{0.7}\right)^{-2/11} \left(\frac{\Omega_{\rm gw}^{95} h^2}{2.3 \times 10^{-10}} \right)^{4/33} \nonumber \\
&\times& \left( \frac{N}{10}\right)^{-4/11} \left(\frac{g_{*}(T_{a_0})}{20} \right)^{1/66} \left( \frac{T_{a_0}}{0.2 \, \textrm{GeV}}  \right)^{28/11}  \nonumber\\
\end{eqnarray}
where $\epsilon_{\rm gw} \simeq 0.7 \pm 0.4$ is an efficiency parameter of the gravitational wave emission \cite{Hiramatsu:2013qaa} , and $\Omega_{\rm gw}^{95}$
is the current 95\% confidence upper limit at $\nu_{\rm 1yr} \simeq 3 \times 10^{-8}$ Hz \cite{Lasky:2015lej}. 
These considerations imply that a small axino coupling ($\sim 1/v_0$) requires correspondingly light axions to be compatible
with the observational data.
In our benchmark parameter choice for the secluded spectrum and thermal yield of axinos,
those constraints are safely satisfied.

\section{Conclusions}
\label{sec:conc}

In this paper, we have studied implications of supersymmetrizing the clockwork axion model.
By supersymmetry, the superpartner axinos have the same clockwork pattern with respect to the coupling hierarchy. 
The coupling hierarchy is not spoiled by SUSY breaking if the SUSY breaking is universal over the clockwork sites. 
Even for non-universal SUSY breaking, the coupling hierarchy is approximately maintained when the SUSY breaking scale
is sufficiently smaller than the clockwork mass scale.
In the universal SUSY breaking case, we find that the clockwork axino mass spectrum can be inverted in ordering when SUSY breaking mass
is larger than the clockwork mass scale. 
The same happens to the saxion sector. 
In this work we have focused on the normal ordering because it may have interesting consequences
for axino dark matter.
Under the assumption that axinos are mainly produced from the SM thermal bath,
 we find that the thermal yield
of the lightest axino is \emph{exponentially} enhanced compared to the non-clockwork axino case with the same
coupling to the SM. 
This is because the lightest axino production is dominated by the decay of heavy axinos
which interact with the SM thermal bath with exponentially larger coupling. 
Thus the relevant parameter space for axino dark matter is significantly different from the conventional non-clockwork axino scenarios.
It generally requires a lower reheating temperature than the conventional scenarios for the same mass of axino dark matter.
Furthermore, we expect that the phase space distribution of the axino dark matter is highly dependent on the detailed clockwork
structure, which may have implications for the structure formation~\cite{bky2020}.
Finally the string-wall network from the superpartner clockwork axions has interesting cosmological consequences on
dark radiation and gravitational waves, imposing an upper bound on the axino coupling for a given clockwork mass scale.  
It may be interesting to examine further cosmological and collider consequences for supersymmetric clockwork models.

\acknowledgments
We thank Jeff Kost and Chang Sub Shin for helpful discussions. 
The work of KJB was supported by
the National Research Foundation of Korea (NRF) grant funded by the Korean
government (NRF-2020R1C1C1012452).
SHI acknowledges support from Basic Science Research Program through 
the National Research Foundation of Korea (NRF)
funded by the Ministry of Education (2019R1I1A1A01060680).
SHI was also supported by IBS under the project code, IBS-R018-D1.

\bibliography{cw_axino}

\end{document}